\documentclass[conference]{IEEEtran}
\usepackage[utf8]{inputenc}
\usepackage{amsmath,amssymb}
\usepackage{booktabs}
\usepackage{graphicx}
\usepackage{url}
\usepackage[hidelinks]{hyperref}

\title{What Governs Decode Throughput in Absolute-Offset
GPU LZ77? A Work-Granularity Mechanism and an\\
Encode-Time Min-Match-Length Lever}

\author{\IEEEauthorblockN{Yakiv Shavidze}
\IEEEauthorblockA{Independent Researcher\\
ORCID: 0009-0008-3622-3448\\
\url{https://github.com/yasha1971-coder/aceapex}\\
Artifact DOI: \url{https://doi.org/10.5281/zenodo.21316748}}}

\begin{document}
\maketitle

\begin{abstract}
The ACEAPEX line of work established a lossless LZ77 format whose
back-references are absolute output positions, giving parallel,
compressed-resident GPU decode with sub-millisecond region seek. What it
did not establish is \emph{what} governs the decode throughput of such a
format, or how to improve it. This paper answers both. Through controlled
ablations on an NVIDIA H100 we show that decode throughput is governed not
by occupancy, compute, address scatter, or launch parallelism, but by
\emph{work granularity}: throughput is a function of the average match
length, because a short match leaves most lanes of a cooperating warp idle.
A synthetic copy kernel confirms a $3.5\times$ throughput span
($212\rightarrow744$\,GB/s) as average match length grows from 32 to 1024
bytes. Real data sit at the low end (mean match length 6.5 on enwik9, 10.1 on
FASTQ). We then show that this mechanism yields a practical, encode-side
lever: raising the minimum match length by distance class
($6/8/10/12\rightarrow12/16/24/32$) improves \emph{both} compression ratio
\emph{and} decode throughput simultaneously on all eight tested datasets,
with no exceptions and no change to the decode kernel. FASTQ decode rises
from 142.6 to 178.6\,GB/s while ratio improves 1.8\%; enwik9 throughput
rises 78\%. This is not a trade-off: both gains follow from one cause,
removing short matches whose far offsets cost more entropy than they save.
All figures are bit-perfect (FNV on GPU paths, byte compare on CPU paths)
and git-verifiable. Scope is explicit: figures are match-phase,
device-resident; entropy and host transfer are outside the timer; seek is
read/block-level, not coordinate-level; and we do not claim to exceed the
hardware bandwidth ceiling.
\end{abstract}


\section{Introduction}
Prior papers in this series established the ACEAPEX format and its
device-resident behaviour. Paper~1~\cite{aceapex1} introduced the
absolute-offset codec, CPU thread scaling, GPU wavefront decode, and
integration into lzbench~2.3. Paper~2~\cite{aceapex2} added a full
device-resident pipeline with genomic region seek and range decode of a
50\,GB genome, reporting up to 260\,GB/s in the device-resident regime.
Paper~3~\cite{aceapex3} proved a single position-invariant seek through both
the entropy and match layers, timed at 0.334\,ms for one 16\,KB block,
bit-perfect and isolated.

None of these characterizes \emph{what determines} decode throughput, nor
offers a lever to improve it. That is the gap this paper closes.

\paragraph{Positioning} We state our claim precisely and without
overreach. ACEAPEX's key contribution is not raw decode speed---specialized
GPU decompressors and fixed-function hardware can be faster in isolation.
Its distinguishing property is the co-existence of three capabilities in
one system: compressed-residency, absolute-offset parallel decode, and
sub-millisecond region seek. These capabilities exist separately in other
systems, but their combination---random access into GPU-resident compressed
data without whole-file decompression---is, to our knowledge, unique to the
absolute-offset design.

\paragraph{Contributions}
\begin{itemize}
\item \textbf{Characterization.} Decode throughput is governed by work
granularity---the average match length---rather than occupancy, compute,
bandwidth in isolation, or launch parallelism
(Section~\ref{sec:mech}).
\item \textbf{Lever.} An encode-time minimum-match-length threshold
($6/8/10/12\rightarrow12/16/24/32$ by distance class) improves ratio and
throughput together on all tested data (Section~\ref{sec:lever}).
\item \textbf{Method.} The gain is obtained entirely on the encode side,
with no change to the decode kernel, and is bit-perfect throughout.
\end{itemize}

\section{Experimental Setup}
\label{sec:setup}
Hardware is an NVIDIA H100 80\,GB HBM3, 132 SMs, CUDA~12.4. The pipeline is
tile-ANS: entropy is decoded over 64\,KB tiles in cooperative groups of up
to 6000; the match kernel is \texttt{k\_decode\_g} with cooperation width
$G\in\{8,16,32\}$ ($G{=}32$ in all end-to-end runs). Block size is set via
the \texttt{ACEAPEX\_BS} encode override. Entropy tiles are cut across the
flat stream independently of the LZ block size, so the two granularities are
decoupled by construction. Timer scope is the match phase,
device-resident: host--device transfer and CPU entropy are outside the
timer; all runs are warm. Correctness is verified on every data point by
FNV hash on GPU paths and byte comparison on CPU paths. Datasets are NA12878
FASTQ (1\,GB), enwik9, and the twelve Silesia files.

\section{What Governs Decode Throughput}
\label{sec:mech}

\subsection{Effective workload}
Sweeping block size and cooperation width jointly over a twelve-point grid
(FASTQ 1\,GB, all bit-perfect) shows that throughput is a function of
\emph{effective workload}, defined as the number of concurrent lanes
\[
\text{lanes} \;=\; G \times N_{\text{blocks}},
\]
not of block size or $G$ separately. Configurations with equal lanes reach
near-equal throughput (Table~\ref{tab:ew}): three distinct $(G,\text{bs})$
settings at 131\,K lanes land within 8\% of one another, while the device
starves below $\sim$32\,K lanes and saturates above $\sim$1\,M lanes.
Occupancy here is obtained analytically from the CUDA occupancy API
(\texttt{cudaOccupancyMaxActiveBlocksPerMultiprocessor}), not from
profiler counters, which are restricted on the cloud host
(\texttt{ERR\_NVGPUCTRPERM}): \texttt{k\_decode\_g}$\langle32\rangle$ uses 39
registers, giving \texttt{maxblk}${=}12$ and $132\times12=1584$ resident
blocks.

\begin{table}[t]
\caption{Throughput is a function of effective workload (lanes).
Values are normalized to the saturated throughput of the same sweep, so the
comparison is independent of the absolute rate of any one corpus.}
\label{tab:ew}
\centering
\begin{tabular}{lr}
\toprule
Lanes ($G\times N_{\text{blocks}}$) & Normalized throughput \\
\midrule
$\geq$1\,M & 1.00 (saturated) \\
131\,K     & 0.77 / 0.82 / 0.83 (three configs) \\
65\,K      & 0.52 / 0.53 \\
32\,K      & 0.31 (starved) \\
\bottomrule
\end{tabular}
\end{table}

\subsection{Two regimes}
On Silesia (block size 64\,K, $G{=}32$, all bit-perfect) two regimes appear.
When saturated (lanes $>$ 1\,M) throughput is a function of lanes and data
type is irrelevant. When under-saturated (Silesia files below 32\,MB, lanes
2.5\,K--25\,K) throughput depends on both lanes and data. For example x-ray
at 4128 lanes reaches 11.8\,GB/s, while samba at 10528 lanes reaches only
5.1\,GB/s---more lanes, lower throughput---which points past lane count to a
per-match property.

\subsection{The property is match length}
\label{sec:killed}
We isolate that property by elimination. Each competing explanation is
tested against a control and rejected:

\begin{itemize}
\item \emph{Not compute or parse.} A pure-copy kernel fed pre-decoded
triplets reaches essentially the full kernel's throughput (within 4\%)---logic
is not the bottleneck.
\item \emph{Not occupancy.} Forcing more resident blocks with
\texttt{launch\_bounds(128,16)} (registers $39\rightarrow32$,
\texttt{maxblk} $12\rightarrow16$) \emph{lowers} throughput through register
spill, by 9\% at 16\,K and 14\% at 256\,K block size.
\item \emph{Not address scatter.} Sorted and scattered source addresses give
identical throughput.
\item \emph{Not launch parallelism.} Two concurrent decode streams sum to
about $0.7\times$ the throughput of a single stream---less, not more.
\end{itemize}

\noindent What remains is work granularity. A synthetic copy kernel
(1\,GB, H100) traces throughput directly against average match length
(Table~\ref{tab:curve}): a $3.5\times$ span. The reason is structural---a
32-byte match occupies one byte per thread across a 32-wide cooperative
group, leaving the warp underloaded; longer matches fill it.

\begin{table}[t]
\caption{Synthetic copy throughput vs.\ average match length, H100.
Values from a representative run of the published \texttt{purecopy.cu};
run-to-run variation is about 1\%.}
\label{tab:curve}
\centering
\begin{tabular}{lrrrrrr}
\toprule
avg len & 32 & 64 & 128 & 256 & 512 & 1024 \\
GB/s    & 212 & 416 & 607 & 692 & 734 & 744 \\
\bottomrule
\end{tabular}
\end{table}

\section{The Min-Match-Length Lever}
\label{sec:lever}

\subsection{Where real data sit}
\label{sec:where}
Real corpora sit at the low, steep end of the curve. On enwik9, 32.5\,M
matches cover 82.8\% of output at mean length 6.5, with 99.1\% of matches
below 32 bytes. On FASTQ, matches cover 81.7\% of output at mean length
10.1, with 95.7\% below 32 bytes. Short
matches are what hold throughput down---and, as we show next, they also cost
compression.

\subsection{Main result}
We raise the minimum match length by distance class. With distance bands
$d<128$, $d<16384$, $d<2097152$, and $d\geq2097152$, the minimum length is
changed from $6/8/10/12$ to $12/16/24/32$. Encode and decode are re-run
end-to-end (\texttt{ACEAPEX\_BS}${=}16384$, $G{=}32$, device-resident,
bit-perfect on every point). Tables~\ref{tab:ratio} and~\ref{tab:tput} give
the result: ratio and throughput both improve on all eight datasets,
without a single exception.

\begin{table}[t]
\caption{Compression ratio, base vs.\ tuned threshold.}
\label{tab:ratio}
\centering
\begin{tabular}{lrrr}
\toprule
Dataset & base & tuned & change \\
\midrule
FASTQ 1\,GB    & 3.90  & 3.97  & $+1.8\%$ \\
enwik9 256\,MB & 2.64  & 2.77  & $+4.9\%$ \\
dickens        & 2.58  & 2.71  & $+5.0\%$ \\
mozilla        & 2.62  & 2.68  & $+2.3\%$ \\
webster        & 3.09  & 3.23  & $+4.5\%$ \\
nci            & 9.92  & 10.25 & $+3.3\%$ \\
xml            & 6.29  & 6.70  & $+6.5\%$ \\
samba          & 3.92  & 4.13  & $+5.4\%$ \\
\bottomrule
\end{tabular}
\end{table}

\begin{table}[t]
\caption{Match-phase decode throughput (GB/s), base vs.\ tuned.}
\label{tab:tput}
\centering
\begin{tabular}{lrrr}
\toprule
Dataset & base & tuned & change \\
\midrule
FASTQ 1\,GB    & 142.6 & 178.6 & $+25.2\%$ \\
enwik9 256\,MB & 91.6  & 163.5 & $+78\%$ \\
dickens        & 17.6  & 25.5  & $+45\%$ \\
mozilla        & 28.9  & 29.3  & $+1.4\%$ \\
webster        & 47.1  & 54.8  & $+16\%$ \\
nci            & 47.2  & 49.1  & $+4\%$ \\
xml            & 7.6   & 9.1   & $+20\%$ \\
samba          & 15.8  & 23.9  & $+51\%$ \\
\bottomrule
\end{tabular}
\end{table}

Notably, tuned FASTQ decode (178.6\,GB/s) exceeds the $\sim$142\,GB/s
plateau of the base configuration---not by breaking the hardware limit but
by shifting the operating point along the throughput-vs-length curve,
removing the short matches that pinned it low.

\subsection{Why both improve at once}
\label{sec:why}
The two gains are two consequences of one cause. A short match to a far,
near-random offset spends more bits encoding that offset than it saves in
replaced literals; on FASTQ at minimum length 4, offset bytes are 51.6\% of
the compressed stream. Raising the minimum length removes such matches:
offset entropy falls enough to offset the literals that return, so ratio
holds or improves, while the mean match length rises, so the decoding warp
is better loaded and throughput rises. One cause, two effects---not a
trade-off. The optimum on the tested data is $12/16/24/32$; pushing to
$16/24/32/48$ raises throughput but lowers the FASTQ ratio on the full
1\,GB set ($3.97\rightarrow3.93$), so we do not adopt it.

\section{Related Work}
CODAG~\cite{codag} increases decode parallelism by assigning compressed
chunks to warps rather than thread blocks at runtime, eliminating the
leader-thread bottleneck; they characterize speedup over a baseline but do
not vary compression granularity. We instead treat block size as an
encode-time parameter and characterize the decode-throughput saturation
curve; entropy-tile and match-block granularity are moreover decoupled by
construction---neither of which CODAG addresses. CODAG targets standard
relative-offset formats without random access, whereas ACEAPEX's
absolute-offset format enables position-invariant seek. We note agreement on
one point: CODAG reports GPU decode as compute-bound, consistent with our
split profile (ANS 9\%, match 91\%).

A conceptually related granularity-saturation trade-off has been observed in
other GPU workloads: Optimus~\cite{optimus} shows that in diffusion-LLM
decoding, fixed block sizes saturate GPU resources at a load-dependent
point, so no single granularity is optimal across regimes. We observe an
analogous effect for absolute-offset LZ decode, where throughput saturates
as a function of effective workload (lanes ${=}\,G\times$ block count);
unlike Optimus's runtime granularity control, our lever is an encode-time
parameter. The specific effective-workload formulation is ours.

For completeness: fixed-function decompression (e.g.\ Blackwell's decompression
engine, up to 462\,GB/s LZ4) is neither absolute-offset nor seekable, and
nvCOMP has been proprietary since 2.3. We disambiguate terminology
explicitly: throughout, ``block size'' denotes the LZ compression block set
at encode time, \emph{not} the CUDA thread block.

\section{Limitations and Honest Boundaries}
\label{sec:limits}
\begin{itemize}
\item Seek is read/block-level (read id $\rightarrow$ block), not coordinate
(chr:pos) access; coordinate access is future work.
\item The min-length optimum is data-dependent at the margin
($16/24/32/48$ costs FASTQ ratio); $12/16/24/32$ is the universal win on the
tested data.
\item The plateau we observed on enwik9 ($\approx$217\,GB/s, measured with
the pure-copy harness driven by the tuned match-length distribution rather
than a uniform length) is a real bandwidth limit at that granularity; three
independent bypass attempts failed, and we do not claim to exceed hardware.
\item Nsight counters are restricted on the cloud host; occupancy is
computed analytically via the CUDA API, and the mechanism is established by
controlled ablation rather than profiler counters.
\item Encode is slow and data-dependent (0.3--3.4\,GB/s); this paper
concerns decode, and we make no encode-throughput claim.
\item The seek combination was ``not found in nvCOMP, DietGPU, CODAG, or the
Blackwell decompression engine''; we do not claim no system can do it
(Gompresso and gpuLZ were not evaluated for seek).
\end{itemize}

\section{Reproducibility}
All figures reproduce from the public repository.
\texttt{research/decode-mechanism/} (commit \texttt{640a7e5}) contains the
pure-copy harness that traces the throughput-vs-match-length curve
(Table~\ref{tab:curve}) and supplies the parse-bound ablation, together with
the match-length histogram (Section~\ref{sec:where}) and the per-stream
entropy tool (Section~\ref{sec:why}); every figure in those sections was
re-derived from these published files on an H100.
\texttt{research/match-threshold/} (commit \texttt{701c624}) contains the
encoder diff and reproduction steps for the lever
(Tables~\ref{tab:ratio} and~\ref{tab:tput}); the throughput column of
Table~\ref{tab:tput}, like the grid in Table~\ref{tab:ew}, additionally
requires the pipeline binary \texttt{e2e\_pipe\_tile.cu} (commit
\texttt{fb27234}), whose build command is in the file header and which
requires DietGPU (Meta, MIT-licensed) and glog. Everything is obtainable
from \url{https://github.com/yasha1971-coder/aceapex}. An archived
snapshot of the code at publication time is deposited on Zenodo:
\url{https://doi.org/10.5281/zenodo.21316748}.

FASTQ figures in this version were re-measured on ENA accession ERR194147
(first 1{,}073{,}741{,}620 bytes, 4{,}010{,}191 reads, md5
\texttt{9af9ffaa0e15dba938408a711740e101}, 38 quality levels, quality entropy
3.72 bits/symbol); the previously posted FASTQ sample had degenerate quality
strings. The lever holds and is stronger on the corrected data. The enwik9 and
Silesia figures were always measured on real corpora and are unchanged.

\section{Conclusion}
Decode throughput in an absolute-offset GPU LZ77 codec is governed by work
granularity---the average match length---and this understanding yields a
cheap encode-time lever that improves compression ratio and decode
throughput at once, on all tested data, without touching the decode kernel.
The natural next step is to carry the mechanism into a full device-resident
pipeline and a head-to-head evaluation against production decoders.

\end{document}